\documentclass[aps,superscriptaddress,prl,showpacs,twocolumn]{revtex4}
\usepackage{graphicx,amssymb,amsmath,subeqnarray} 
\newcommand{\Nt}{N}
\newcommand{\rhob}{N}
\newcommand{\rholb}{\tilde N}

\begin{document}
\title{Traffic jams and intermittent flows in microfluidic networks}
\author{Nicolas Champagne}
\author{Romain Vasseur}
\author{Adrien Montourcy}
\author{Denis Bartolo}
\affiliation{PMMH, CNRS, ESPCI ParisTech, Universit\'e Paris 6, Universit\'e Paris 7, \\10, rue Vauquelin, 75231 Paris cedex 05 FRANCE}
\begin{abstract}
We investigate both experimentally and theoretically the traffic of particles flowing in  microfluidic obstacle networks. We show that the traffic dynamics is a non-linear process: the particle current does not scale with the particle density even in the dilute limit where no particle collision occurs. We  demonstrate  that this non-linear behavior stems from  long range hydrodynamic interactions. 
Importantly, we also establish that there exists a maximal current above which no stationary particle flow can be sustained. For higher current values, intermittent traffic jams form thereby inducing the ejection of the particles from the initial path and the subsequent invasion of the network. 
Eventually, we  put our findings in the  broader context of  the transport proccesses of driven particles in low dimension.
\end{abstract}
\pacs{47.61.Fg, 47.61.Jd, 47.56.+r}
\maketitle
Hundreds of industrial and biological processes ultimately rely on the transport of particle or droplet suspensions in channels or obstacle networks. Prominent examples include:  particle filtration~\cite{cazabat1990,roussel2007}, size separation of colloids and polymers~\cite{heftmann1983,huang2004}, emulsion flows through porous rocks during enhanced oil recovery, droplet transport in microfluidic chips~\cite{joanicot2005, teh2008} and,  obviously, blood microflows~\cite{popel2005}. For all these examples, the particle size typically compares with the distance between the obstacles, or walls, forming the network. Therefore, even in the dilute limit, the particles locally reduce the network conductivity, thereby inducing a dynamic redistribution of the fluid flow. This non-trivial interplay between the position of the advected particles and the driving flow field makes the description of the particle traffic a challenging task. Whereas all the above examples typically involve very large networks, most of the recent investigations of such "dynamic" networks have been hitherto focused on droplet traffic in minimal microfluidic set-ups composed of a single loop or of a single T-junction. Nonetheless, despite the apparent simplicity of these geometries, a rich variety of periodic and aperiodic drop dynamics have been reported  due to finite time anti-correlations of the droplets trajectories at the junctions~\cite{engl2005,jousse2005,jousse2007,fuerstman2007,schindler2008,belloul2009,sessoms2009}. 

\begin{figure}[b]
\begin{center}
\includegraphics[width=\columnwidth]{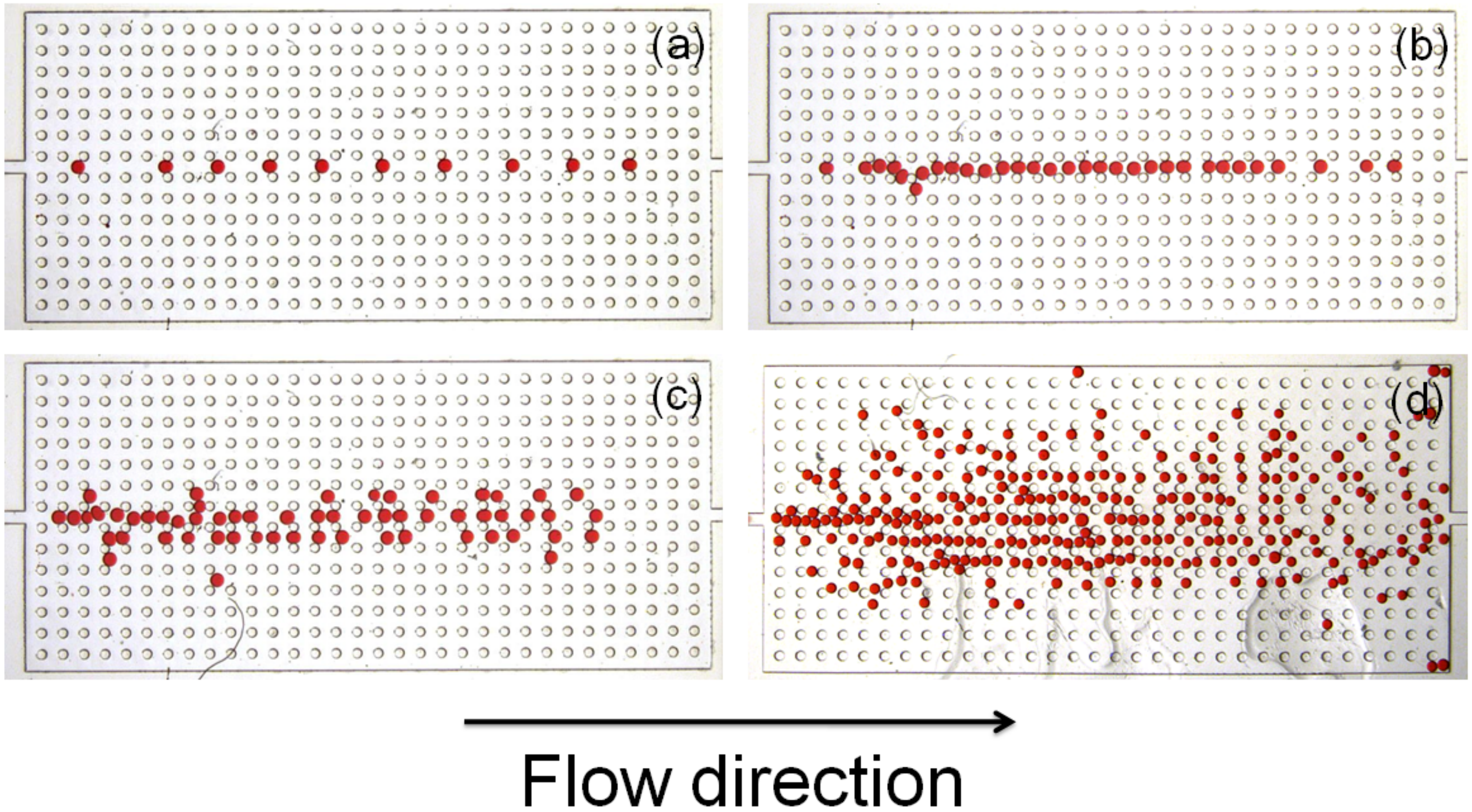}
\caption{Water droplets in hexadecane oil flowing in a square lattice of obstacles. (a) Imposed droplet current: $j=0.4$ Hz. (b) Droplet current: $j=0.7$ Hz. A jam forms at the entrance of the network. (c) Droplet current: $j=2$ Hz. (d) Very high current phase. The droplets invade the whole network.  Length of the network: 9.75 mm.}
\label{fig1}
\end{center}
\end{figure}

In this letter, we aim at providing a generic description of the trafficking dynamics  in large fluidic networks. More precisely, we focus on a wide class of prototypal microfluidic  networks where a single particle would follow a unique deterministic trajectory connecting a single inlet to a single outlet, see Fig.~\ref{fig1}a.  Increasing the particle injection rate, $j$,  we observed that above a critical value $j^*$, a traffic jam builds up at the inlet. This local increase of the density results in the ejection of the particles, which then invade a larger  fraction of the network and explore alternative exit paths, Fig.~\ref{fig1}b,
c,  and d. Surprisingly, this invasion transition occurs  at a very low apparent volume fraction for which the particles do not experience any collision. 
In the high current regime, we also report on the collective ejection of the particles out of the initial lane which results in an intermittent traffic flow.
We combine experimental investigations with a minimal theoretical model to uncover the physical origin of the jamming and of the invasion dynamics of fluidic networks.  

Our experiment consists in flowing periodic sequences of droplets  through an obstacle network inside a microfluidic channel. The network is made
 of a 15$\times$33 square lattice of cylindrical posts, Figure~\ref{fig1}. The diameter of the posts is $150\,\mu$m, the lattice spacing, $\ell$, is $150\,\mu$m and the channel height is $80\,\mu$m.   Changing the lattice size and the post shape did not yield any qualitatively different results (results not shown). 
We chose this design so that  a single droplet follows a unique deterministic trajectory along the central lane, which is the shortest path connecting the entrance to the exit. 
\begin{figure}[t]
\begin{center}
\includegraphics[height=0.6\columnwidth]{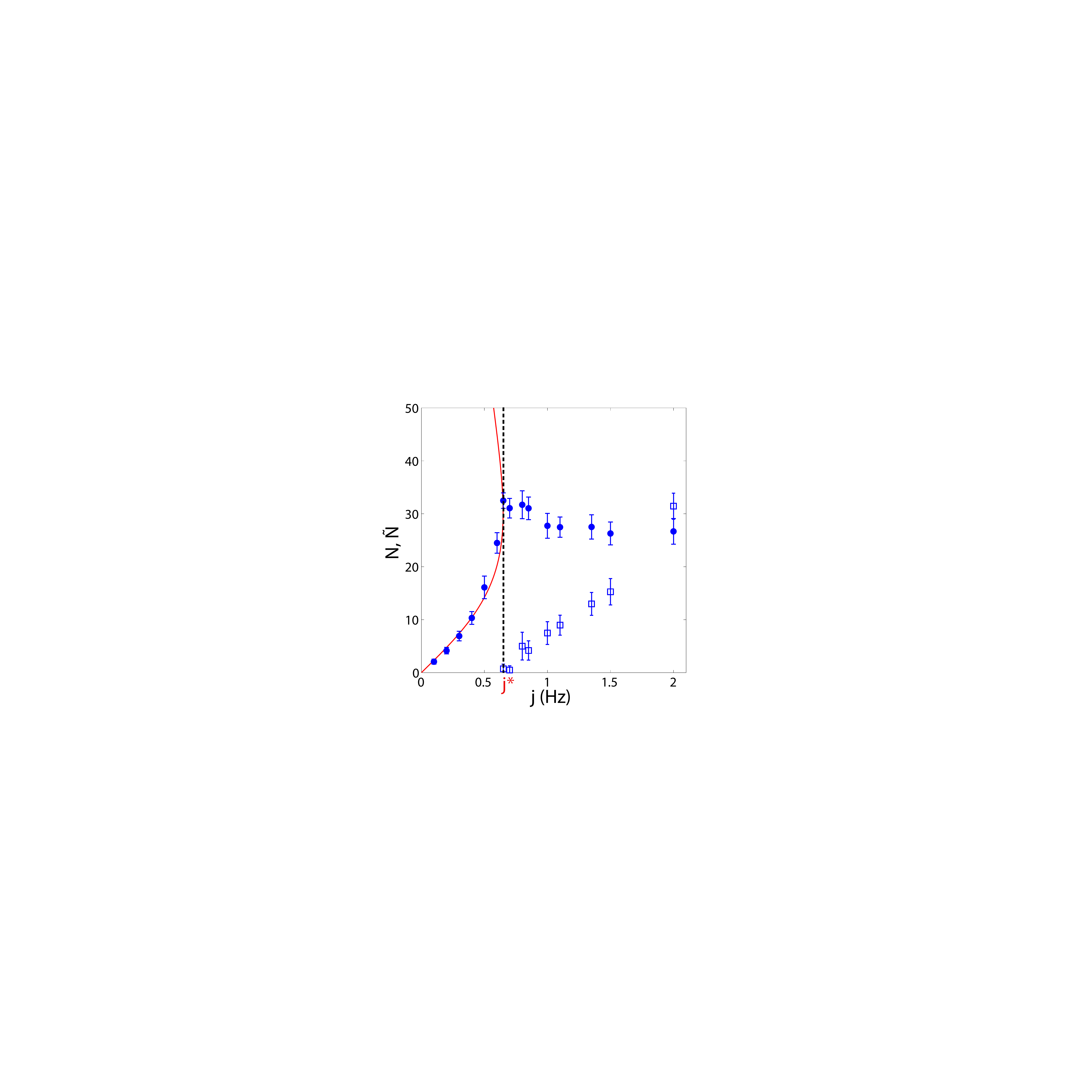}
\caption{ Number of droplets in the central lane: $N$ (\textbullet),   and  number of droplets in the other lateral lanes, $\rholb$, ($\Box$) plotted versus $j$ (experiments).  Full red line: theoretical prediction for the variations of  $\rhob$ according to Eq.~\ref{central} (best fit). The dashed vertical line indicates where the invasion transition occurs: $j=j^*$.}
\label{fig2}
\end{center}
\end{figure}
We used  water droplets (viscosity $\eta=$1 mPa.s) as the advected particles rather than solid microbeads to avoid irreversible clogging of the network.    The droplets were made and transported in  a mixture of hexadecane (viscosity $\eta=2$mPa.s) and span 80 surfactant (3wt$\%$). To produce the droplets {\it in situ} we used a home made drop-on-demand device introduced previously in~\cite{galas2009}. 
This device allowed us to control independently and accurately: the droplet size, the flow rate of the continuous phase and the droplet current:~$j$, i.e. the number of droplets per second injected in the device. $j$ was varied over 2 decades from 0.05 to 5 Hz. In order to ensure a long term stability and a fast dynamic response of the droplet emitter, we used a microfluidic sticker made of a photocurable optical adhesive (NOA81, Norland Products) for the main microfluidic channel~\cite{bartolo2008}. In all the experiments, we used droplets with a radius $a=89\pm4\,\mu$m and  the oil flow rate was set so that the velocity of a single droplet in the center of the channel was $v_{\rm D}^0=350$ $\mu$m$/$s. Thus, both the Reynolds number, $Re\sim10^{-3}$ and the capillary number, $Ca\sim10^{-4}$  were fixed at a constant small value in all our experiments.

\begin{figure}[t]
\begin{center}
\includegraphics[height=0.6\columnwidth]{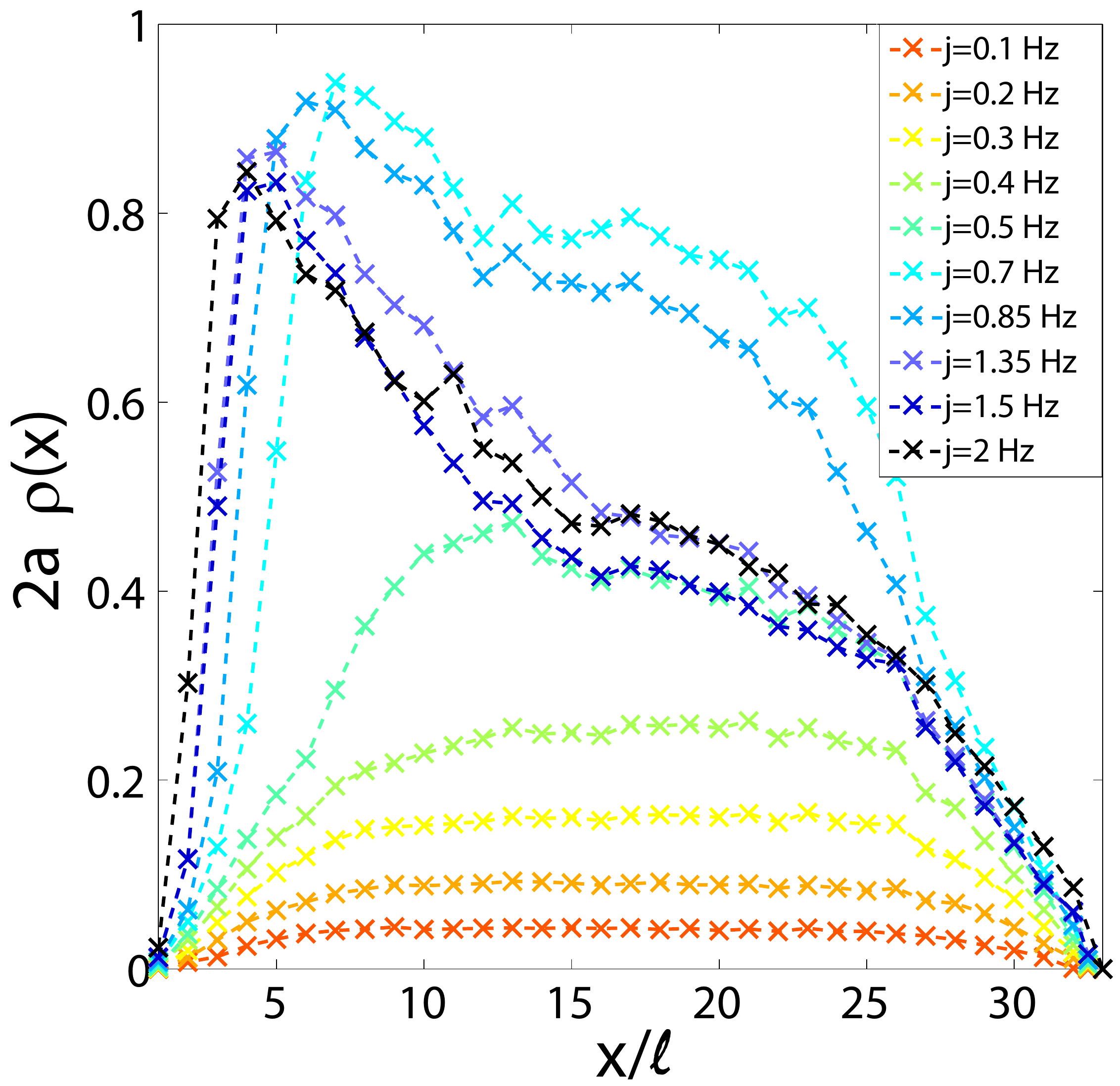}
\caption{Time averaged density profiles normalized by the close packing density $1/(2a)$ in the central lane for different imposed droplet currents $j$. The position is normalized by the lattice spacing, $\ell$. }
\label{fig3}
\end{center}
\end{figure}

We first show in Figure 2, how the numbers of droplets in the central lane, $N$, and outside the central lane, $\rholb$, change as the imposed droplet current $j$ increases. We also plot in Fig.~\ref{fig3} the density field  in the central lane, $\rho(x)$, normalized by its maximal close-packing value, $1/(2a)$, where again $a$ is the droplet diameter.  $\rho(x)$, $\rhob$ and $\rholb$, are time-averaged quantities  systematically measured over 25-minute experiments, performed 10 minutes after the injections of the first droplets at a given current $j$. 

 In the very small current limit, $j\lesssim0.3$ Hz,  $\rhob(j)$  increases linearly with $j$, Fig. 2. The particles keep on following the central lane as an isolated particle would do. The local density $\rho(x)$ simply reflects the $x$ component of the velocity field, $v_{\rm F}^0(x)$, of the continuous phase in a particle free network,~Fig.~\ref{fig3}. The  imposed droplet current, and  the local density are linearly related: $j\propto\rho(x)v_{\rm F}^0(x)$. 
 The variations of $\rho$ close to the entrance and the exit are due to the compression (resp. elongation) of the bare flow field at the entrance (resp. exit) of the network. In all that follows, we  ignore this linear hydrodynamic effect and focus only on the bulk properties of the $\rho$ field.
Increasing the current, the traffic flow crosses-over to a qualitatively different regime. 

For $0.3$ Hz$\lesssim j\leq0.6$ Hz, the droplets still follow the central lane and $\rho(x)$ retains the same flat profile, Fig.~\ref{fig3}. However, the $N(j)$ curve significantly deviates from its initial linear behavior and its variations diverge as the current approaches a well-defined "critical value" $j^*=0.6$ Hz, see Fig.~\ref{fig2}. In fact, these non-linear variations reflect the slowing down of the droplets as their linear density increases. We now stress on one of our most important result: $j^*$ is the maximal stationary particle current that can be sustained in the network.
Indeed, as  $j$ reaches $j^*$, the uniform and stationary droplet distribution becomes unstable: a jam  forms a the entrance of the network, the droplets escape the central lane and invade the network, Figs.~\ref{fig1} and~\ref{fig2}.
Surprisingly, this instability takes place though the droplet density remains much smaller than the close-packing density, $\rho(x)<0.7/(2a)$, and no droplet collision occurs. We can thus unambiguously infer that long range hydrodynamic interactions are shaping the non-linear $(j,N)$ constitutive relation in Fig.~\ref{fig2} and are  responsible for the unexpected destabilization of the traffic flow at $j^*$. Before describing the higher current regimes, we now introduce a theoretical model to uncover the origin of these effective hydrodynamic interactions.

 Since we investigate rather dilute systems, we use a continuous description that substantially simplifies the algebra. This simplification is done without loss of generality in the far field limit. Firstly, we model the obstacle network by a lattice of 1D channels solely characterized by  their local hydrodynamic conductivity $G$.  All the specifics of the obstacle shape are hidden in this parameter which relates the local pressure gradient to the local fluid velocity: ${\bf v_{\rm F}({\bf r})}\equiv -G({\bf r}) \nabla P(\bf r)$, where ${\bf r}$ is the position of the considered elementary channel in the 2D network. Using the incompressibility relation, $\nabla \cdot {\bf v}_{\rm F}=0$, we then obtain the equation for the pressure field:
\begin{equation}
\nabla\cdot  \left [  G({\bf r})\nabla P({\bf r})\right]=0,
\label{laplace}
\end{equation}  
with the boundary condition: $\nabla P=-(v_{\rm F}/G)\hat x$, where $v_{\rm F}$ is the constant fluid velocity far from the particles. 
For a regular lattice of obstacles, $G$ is constant. Secondly, following Jousse and coworkers~\cite{jousse2005,jousse2007} we describe the advected droplets as constant pointwise perturbations to the local conductance. Indeed, the droplet size is smaller than the lattice spacing, $\ell$, which defines the cut-off of our continuous theory. More precisely, for $N$ particles located at ${\bf{r}}_i(t)=(x_i,0)$:
\begin{equation}
G({\bf r})=G\left[1-b^2\sum_{i=1}^N\delta({\bf r-r_i}(t))\right],
\label{conductance}
\end{equation}
where the $\delta$ function is regularized at short distance: $b^2\delta(0)\sim(b/\ell)^2$.  Here $b/\ell$ measures  how much a droplet locally hinders the fluid flow.  Thirdly, the last ingredient of this model is the droplet advection rule. For sake of simplicity, we assume that a droplet located at ${\bf r}_i(t)$ moves with a velocity $\dot {\bf r}_i=\mu {\bf v}_{F}({\bf r}_i)$ where $\mu$ is a constant mobility coefficient.  As the flow is potential, we can now take advantage of an obvious electrostatic analogy. Indeed, for $\Nt=1$, Eqs.~\ref{laplace} and~\ref{conductance}  correspond to the equation for the electric potential induced by a pointwise dielectric particle polarized by a homogeneous electric field~\cite{jackson}. Therefore, we readily deduce that, in the far field limit, a single droplet acts as a source dipole oriented along the $x$-axis.  More quantitatively, we can compute the amplitude of the  corresponding flow disturbance  along $x$ at ${\bf r}=(x,0)$ by solving perturbatively Eq.~\ref{laplace}. At first order in $b^2$: $v_{\rm dip}(x;\dot x_i)=\dot x_i b^2(1-\mu)/(2\pi\mu)\partial_x\frac{1}{|x-x_i|}$. We stress on the sign of this perturbation,  which is opposed to the driving flow along the $x$-axis. Therefore,  it reduces the velocity of two approaching droplets whatever their relative position.
We now write down an effective equation for the dynamics of  $N$ aligned particles. To do so we exploit the pairwise additivity of the hydrodynamic interactions  within the above dipolar approximation. Hence, the equations of motion take the simple form: 
$
\partial_tx_i(t)={v}_{\rm D}^0+\mu\sum_{j\neq i} v_{\rm dip}(x_j-x_i;\dot x_j),
$
where $v_{\rm D}^0$ is the velocity of a single advected droplet.
These $N$ equations can be solved analytically when the droplets are equally separated along a single lane by a distance $L/N$, where $L$ is the system size, and all moving at the same speed, $\dot{x}_i=v_{\rm D}$. This corresponds to  the only stationary solution we observed in our experiments.
Summing over all the dipolar interactions is straightforward and yields a non-linear relation between  the density $\rho=N/L$ and the uniform  particle current: $j=\rho v_{\rm D}$. In the large $N$ limit:
\begin{equation}
j= \frac{\rho v_{\rm D}^0}{1+\frac{\pi}{6}(1-\mu) b^2\rho^2}
\label{central}
\end{equation}
As we measured the velocity $v_{D}^0$ independently , we are left with a single fitting parameter. For $(1-\mu)b^2=0.17$~mm$^{-2}$
,  our first order prediction is in excellent agreement with our experimental data, see the red curve in Fig.~\ref{fig2}.  Our second central theoretical result is that  the the $j(\rho)$ function defined in Eq. 3 has an absolute maximum at $j=j^*=0.65$ Hz, implying that there cannot exist any stationary solution for $j>j^*$. This result  provides a quantitative explanation for the traffic destabilization observed experimentally above this specific current value.



We  shall add  that the above model is not restricted to 2D square lattices. It is indeed straightforward to extend it to any homogeneous potential Stokes flow, which encompasses flows in bidimensional and tridimensional regular networks of arbitrary symmetry and droplet transport in Hele-Shaw geometries. For instance, our prediction is in agreement and generalize the hydrodynamic slowing down of droplets in obstacle-free shallow channels reported  in
~\cite{beatus2006,beatus2007}.

\begin{figure}
\begin{center}
\includegraphics[width=0.8\columnwidth]{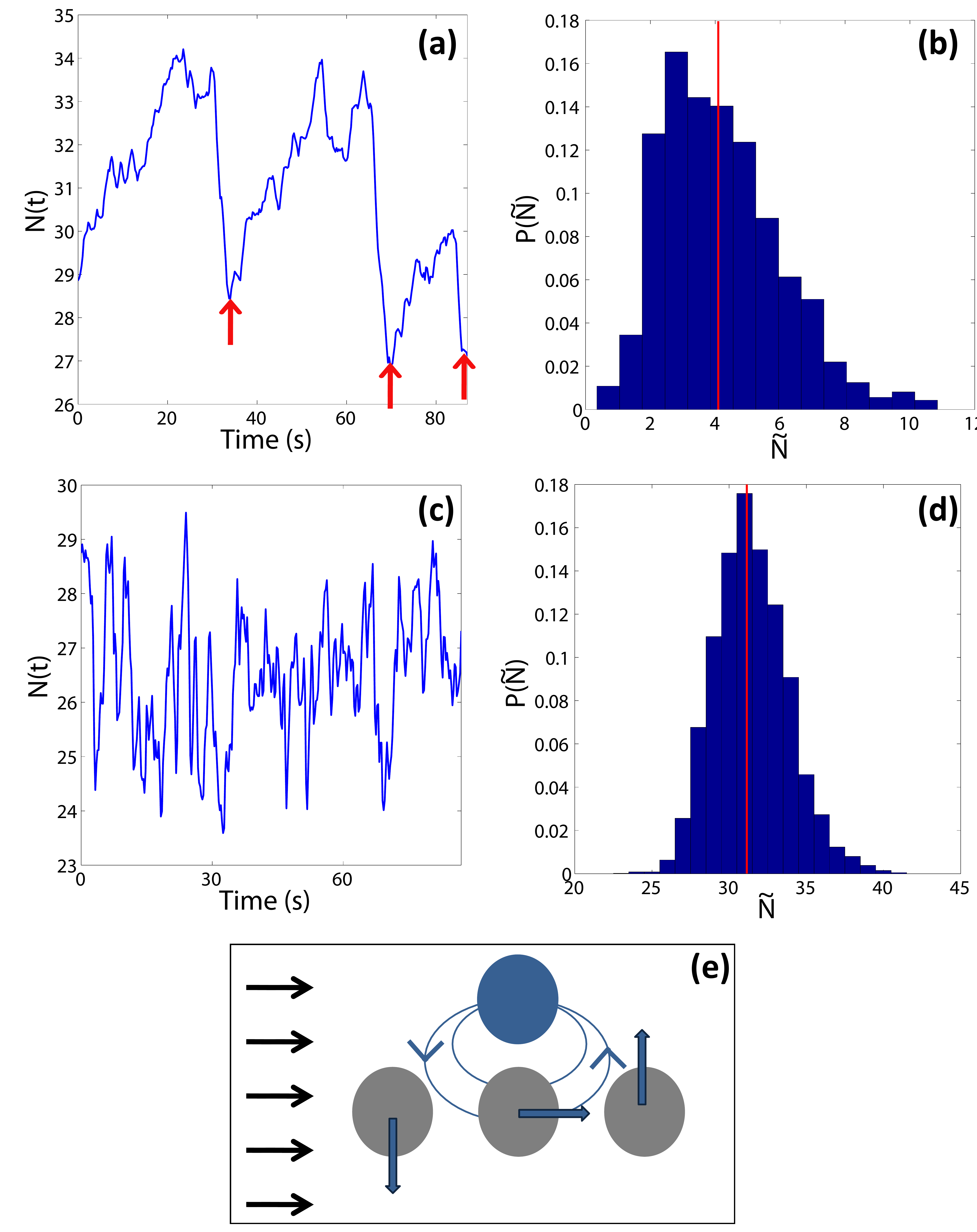}
\caption{(a) Variations of $N(t)$ in the central lane for $j=0.85$ Hz. The red arrow indicates cooperative-ejection events.(b) Histogram of the number of droplets flowing outside the central lane $P(\tilde N)$, computed for 1050 droplets injected in the network at a rate $j=0.85$ Hz. Vertical line: mean value (c) $N(t)$ curve for $j=1.2$ Hz, same time interval as in (a).  (d) Histogram $P(\tilde N)$, computed for 3000 droplets injected in the network at a rate $j=1.2$ Hz.  Sketch of the effective hydrodynamic interactions between 3 advected droplets flowing on two adjacent lanes.}
\label{fig4}
\end{center}
\end{figure}

In the last part of this letter, we focus on the  traffic dynamics in the high current regime $j>j^*$.  For current values larger but close to $j^*$, typically for $j^*<j\lesssim 1$ Hz,  a jam forms at the entrance of the microfluidic network: the instantaneous density field $\rho(x,t)$ becomes asymmetric and its maximum value continuously increases until the droplets locally contact each other, Figs.~\ref{fig1} and~\ref{fig3}. Subsequently, droplets are ejected toward the upper and lower lanes and start invading the network, Figs.~\ref{fig1} and~\ref{fig2}. Consequently, the local density in the central lane is reduced but it does not reach a stationary value, as the jam reforms slowly and disrupts again quickly:  the traffic flow becomes intermittent.  To gain a better insight on this invasion dynamics, we plot in Fig.~\ref{fig4} a, the variations of the instantaneous number of droplets in the central lane, $N(t)$. We clearly distinguish two time scales in this asymmetric signal: the slow one corresponds to the jam formation, where $N(t)$ increases, and the faster one   corresponds to the time during which droplets are cooperatively ejected from their initial path and $N(t)$ decreases. In addition, we plot in Fig.~\ref{fig4} b, the probability distribution,  $P(\tilde N)$, of the number of particles traveling outside the central lane for $j=0.85$ Hz. This distribution is clearly non-Gaussian, which is another signature of the intermittent dynamics and of the correlated ejection of groups of droplets. Rare events, during which groups of particles travel cooperatively through the lateral lanes, broaden the distribution and increase the mean number of particles ejected out of the central lane above its typical value.  
The cooperative ejection of  the droplets can be qualitatively understood by looking at the symmetry of the dipolar flow field. As sketched in Fig.~\ref{fig4}, the droplets moving above and below  the central lane increase the transverse flow downstream, thereby inducing the pumping of the droplets from the central to the side lanes, see the picture in Fig~\ref{fig4}c. In turn, this pumping mechanism is amplified as the number of droplets on side lanes increases, which results in an unstable dynamics yielding the correlated ejection of several droplets when the local density reaches a sufficiently high value.

Finally we briefly describe the high current regime: $j>1$ Hz. For imposed currents much larger than $j^*$, the traffic flow remains non-stationary. However, the jams forms and disrupts continuously without any clear separation between the time scale of these two processes, see Fig.~\ref{fig4} c. Consequently, the distribution of $\tilde N$ becomes more symmetric, and seems to converge toward a Gaussian behavior in the large $j$ limit, Fig.~\ref{fig4} d. In addition, we found that above $j^*$, the system self-organizes so that the density field along the initially preferred path converges and becomes independent of the injected current value. To go beyond this first attempt at understanding the traffic dynamics above $j^*$, a more thorough statistical analysis would be required. Work along this line is in progress.

In summary, we have demonstrated that when particles are advected in a fluidic network, they locally reduce the network conductance  thereby inducing non-local dipolar interactions between the moving objects. The main consequence of this hydrodynamic coupling is the limitation of the maximal stationary current, $j^*$ that can be sustained in the network. Above $j^*$, the initially preferred path becomes unstable and the particles intermittently invade the whole network.  To close this letter, we note that the destabilization and the jamming of stationary traffic flows go beyond the present study. The existence of a maximal current is indeed a very generic feature reported for numerous classes of theoretical models of  driven interacting-particle systems, including discrete exclusion processes~\cite{derridabook}  and phenomenological continuous models of self-driven particles (pedestrian or vehicles)~\cite{helbing2001}. All these models possess a common feature: a non-monotonic constitutive relation relating the current to the particle density . To our knowledge, we provide the first experimental evidence of such a maximal current. 

We thank Julien Tailleur for insightful comments and suggestions. Raphael Voituriez is also ackowledged for an illuminating discussion. 

\begin{thebibliography}{21}
\expandafter\ifx\csname natexlab\endcsname\relax\def\natexlab#1{#1}\fi
\expandafter\ifx\csname bibnamefont\endcsname\relax
  \def\bibnamefont#1{#1}\fi
\expandafter\ifx\csname bibfnamefont\endcsname\relax
  \def\bibfnamefont#1{#1}\fi
\expandafter\ifx\csname citenamefont\endcsname\relax
  \def\citenamefont#1{#1}\fi
\expandafter\ifx\csname url\endcsname\relax
  \def\url#1{\texttt{#1}}\fi
\expandafter\ifx\csname urlprefix\endcsname\relax\def\urlprefix{URL }\fi
\providecommand{\bibinfo}[2]{#2}
\providecommand{\eprint}[2][]{\url{#2}}

\bibitem[{\citenamefont{Hulin et~al.}(1990)\citenamefont{Hulin, Cazabat, Guyon,
  and Carmona}}]{cazabat1990}
\bibinfo{author}{\bibfnamefont{J.-P.} \bibnamefont{Hulin}},
  \bibinfo{author}{\bibfnamefont{A.-M.} \bibnamefont{Cazabat}},
  \bibinfo{author}{\bibfnamefont{E.}~\bibnamefont{Guyon}}, \bibnamefont{and}
  \bibinfo{author}{\bibfnamefont{F.}~\bibnamefont{Carmona}},
  \emph{\bibinfo{title}{Hydrodynamics of Dispersed Media}}
  (\bibinfo{publisher}{Elsevier}, \bibinfo{year}{1990}).

\bibitem[{\citenamefont{Roussel et~al.}(2007)\citenamefont{Roussel, Nguyen, and
  Coussot}}]{roussel2007}
\bibinfo{author}{\bibfnamefont{N.}~\bibnamefont{Roussel}},
  \bibinfo{author}{\bibfnamefont{T.~L.~H.} \bibnamefont{Nguyen}},
  \bibnamefont{and} \bibinfo{author}{\bibfnamefont{P.}~\bibnamefont{Coussot}},
  \bibinfo{journal}{Phys. Rev. Lett.} \textbf{\bibinfo{volume}{98}},
  \bibinfo{pages}{114502} (\bibinfo{year}{2007}).

\bibitem[{\citenamefont{E.Heftmann}(1983)}]{heftmann1983}
\bibinfo{author}{\bibnamefont{E.Heftmann}},
  \emph{\bibinfo{title}{Chromatography:Fundamentals and Applications of
  Chromatographic and Electrophoretic Methods}} (\bibinfo{publisher}{Elsevier
  Science}, \bibinfo{year}{1983}).

\bibitem[{\citenamefont{Huang et~al.}(2004)\citenamefont{Huang, Cox, Austin,
  and Sturm}}]{huang2004}
\bibinfo{author}{\bibfnamefont{L.~R.} \bibnamefont{Huang}},
  \bibinfo{author}{\bibfnamefont{E.~C.} \bibnamefont{Cox}},
  \bibinfo{author}{\bibfnamefont{R.~H.} \bibnamefont{Austin}},
  \bibnamefont{and} \bibinfo{author}{\bibfnamefont{J.~C.} \bibnamefont{Sturm}},
  \bibinfo{journal}{Science} \textbf{\bibinfo{volume}{304}},
  \bibinfo{pages}{987} (\bibinfo{year}{2004}).

\bibitem[{\citenamefont{Joanicot and Ajdari}(2005)}]{joanicot2005}
\bibinfo{author}{\bibfnamefont{M.}~\bibnamefont{Joanicot}} \bibnamefont{and}
  \bibinfo{author}{\bibfnamefont{A.}~\bibnamefont{Ajdari}},
  \bibinfo{journal}{Science} \textbf{\bibinfo{volume}{309}},
  \bibinfo{pages}{887} (\bibinfo{year}{2005}).

\bibitem[{\citenamefont{Teh et~al.}(2008)\citenamefont{Teh, Lin, Hung, and
  Lee}}]{teh2008}
\bibinfo{author}{\bibfnamefont{S.-Y.} \bibnamefont{Teh}},
  \bibinfo{author}{\bibfnamefont{R.}~\bibnamefont{Lin}},
  \bibinfo{author}{\bibfnamefont{L.-H.} \bibnamefont{Hung}}, \bibnamefont{and}
  \bibinfo{author}{\bibfnamefont{A.~P.} \bibnamefont{Lee}},
  \bibinfo{journal}{Lab Chip} \textbf{\bibinfo{volume}{8}},
  \bibinfo{pages}{198} (\bibinfo{year}{2008}).

\bibitem[{\citenamefont{Popel and Johnson}(2005)}]{popel2005}
\bibinfo{author}{\bibfnamefont{A.~S.} \bibnamefont{Popel}} \bibnamefont{and}
  \bibinfo{author}{\bibfnamefont{P.~C.} \bibnamefont{Johnson}},
  \bibinfo{journal}{Annual Review of Fluid Mechanics}
  \textbf{\bibinfo{volume}{37}}, \bibinfo{pages}{43} (\bibinfo{year}{2005}).

\bibitem[{\citenamefont{Engl et~al.}(2005)\citenamefont{Engl, Roche, Colin,
  Panizza, and Ajdari}}]{engl2005}
\bibinfo{author}{\bibfnamefont{W.}~\bibnamefont{Engl}},
  \bibinfo{author}{\bibfnamefont{M.}~\bibnamefont{Roche}},
  \bibinfo{author}{\bibfnamefont{A.}~\bibnamefont{Colin}},
  \bibinfo{author}{\bibfnamefont{P.}~\bibnamefont{Panizza}}, \bibnamefont{and}
  \bibinfo{author}{\bibfnamefont{A.}~\bibnamefont{Ajdari}},
  \bibinfo{journal}{Phys. Rev. Lett.} \textbf{\bibinfo{volume}{95}},
  \bibinfo{pages}{208304} (\bibinfo{year}{2005}).

\bibitem[{\citenamefont{Jousse et~al.}(2005)\citenamefont{Jousse, Lian, Janes,
  and Melrose}}]{jousse2005}
\bibinfo{author}{\bibfnamefont{F.}~\bibnamefont{Jousse}},
  \bibinfo{author}{\bibfnamefont{G.}~\bibnamefont{Lian}},
  \bibinfo{author}{\bibfnamefont{R.}~\bibnamefont{Janes}}, \bibnamefont{and}
  \bibinfo{author}{\bibfnamefont{J.}~\bibnamefont{Melrose}},
  \bibinfo{journal}{Lab Chip} \textbf{\bibinfo{volume}{5}},
  \bibinfo{pages}{646} (\bibinfo{year}{2005}).

\bibitem[{\citenamefont{Jousse et~al.}(2006)\citenamefont{Jousse, Farr, Link,
  Fuerstman, and Garstecki}}]{jousse2007}
\bibinfo{author}{\bibfnamefont{F.}~\bibnamefont{Jousse}},
  \bibinfo{author}{\bibfnamefont{F.}~\bibnamefont{Farr}},
  \bibinfo{author}{\bibfnamefont{D.~R.} \bibnamefont{Link}},
  \bibinfo{author}{\bibfnamefont{M.~J.} \bibnamefont{Fuerstman}},
  \bibnamefont{and}
  \bibinfo{author}{\bibfnamefont{P.}~\bibnamefont{Garstecki}},
  \bibinfo{journal}{Phys. Rev. E} \textbf{\bibinfo{volume}{74}},
  \bibinfo{pages}{036311} (\bibinfo{year}{2006}).

\bibitem[{\citenamefont{Schindler and Ajdari}(2008)}]{schindler2008}
\bibinfo{author}{\bibfnamefont{M.}~\bibnamefont{Schindler}} \bibnamefont{and}
  \bibinfo{author}{\bibfnamefont{A.}~\bibnamefont{Ajdari}},
  \bibinfo{journal}{Phys. Rev. Lett.} \textbf{\bibinfo{volume}{100}},
  \bibinfo{pages}{044501} (\bibinfo{year}{2008}).

\bibitem[{\citenamefont{Belloul et~al.}(2009)\citenamefont{Belloul, Engl,
  Colin, Panizza, and Ajdari}}]{belloul2009}
\bibinfo{author}{\bibfnamefont{M.}~\bibnamefont{Belloul}},
  \bibinfo{author}{\bibfnamefont{W.}~\bibnamefont{Engl}},
  \bibinfo{author}{\bibfnamefont{A.}~\bibnamefont{Colin}},
  \bibinfo{author}{\bibfnamefont{P.}~\bibnamefont{Panizza}}, \bibnamefont{and}
  \bibinfo{author}{\bibfnamefont{A.}~\bibnamefont{Ajdari}},
  \bibinfo{journal}{Phys. Rev. Lett.} \textbf{\bibinfo{volume}{102}},
  \bibinfo{pages}{194502} (\bibinfo{year}{2009}).

\bibitem[{\citenamefont{Sessoms et~al.}(2009)\citenamefont{Sessoms, Belloul,
  Engl, Roche, Courbin, and Panizza}}]{sessoms2009}
\bibinfo{author}{\bibfnamefont{D.~A.} \bibnamefont{Sessoms}},
  \bibinfo{author}{\bibfnamefont{M.}~\bibnamefont{Belloul}},
  \bibinfo{author}{\bibfnamefont{W.}~\bibnamefont{Engl}},
  \bibinfo{author}{\bibfnamefont{M.}~\bibnamefont{Roche}},
  \bibinfo{author}{\bibfnamefont{L.}~\bibnamefont{Courbin}}, \bibnamefont{and}
  \bibinfo{author}{\bibfnamefont{P.}~\bibnamefont{Panizza}},
  \bibinfo{journal}{Phys. Rev. E} \textbf{\bibinfo{volume}{80}},
  \bibinfo{pages}{016317} (\bibinfo{year}{2009}).

\bibitem[{\citenamefont{Fuerstman et~al.}(2007)\citenamefont{Fuerstman,
  Garstecki, and Whitesides}}]{fuerstman2007}
\bibinfo{author}{\bibfnamefont{M.~J.} \bibnamefont{Fuerstman}},
  \bibinfo{author}{\bibfnamefont{P.}~\bibnamefont{Garstecki}},
  \bibnamefont{and} \bibinfo{author}{\bibfnamefont{G.~M.}
  \bibnamefont{Whitesides}}, \bibinfo{journal}{Science}
  \textbf{\bibinfo{volume}{315}}, \bibinfo{pages}{828} (\bibinfo{year}{2007}).

\bibitem[{\citenamefont{Galas et~al.}(2009)\citenamefont{Galas, Bartolo, and
  Studer}}]{galas2009}
\bibinfo{author}{\bibfnamefont{J.-C.} \bibnamefont{Galas}},
  \bibinfo{author}{\bibfnamefont{D.}~\bibnamefont{Bartolo}}, \bibnamefont{and}
  \bibinfo{author}{\bibfnamefont{V.}~\bibnamefont{Studer}},
  \bibinfo{journal}{New Journal of Physics} \textbf{\bibinfo{volume}{11}},
  \bibinfo{pages}{075027 (11pp)} (\bibinfo{year}{2009}).

\bibitem[{\citenamefont{Denis~Bartolo and Studer}(2008)}]{bartolo2008}
\bibinfo{author}{\bibfnamefont{P.~N.} \bibnamefont{Denis~Bartolo},
  \bibfnamefont{Guillaume~Degr\'e}} \bibnamefont{and}
  \bibinfo{author}{\bibfnamefont{V.}~\bibnamefont{Studer}},
  \bibinfo{journal}{Lab Chip} \textbf{\bibinfo{volume}{8}},
  \bibinfo{pages}{274} (\bibinfo{year}{2008}).

\bibitem[{\citenamefont{Jackson}(1998)}]{jackson}
\bibinfo{author}{\bibfnamefont{J.~D.} \bibnamefont{Jackson}},
  \emph{\bibinfo{title}{Classical electrodynamics}} (\bibinfo{publisher}{John
  Wiley and Sons}, \bibinfo{year}{1998}).

\bibitem[{\citenamefont{Beatus et~al.}(2006)\citenamefont{Beatus, Tlusty, and
  Bar-Ziv}}]{beatus2006}
\bibinfo{author}{\bibfnamefont{T.}~\bibnamefont{Beatus}},
  \bibinfo{author}{\bibfnamefont{T.}~\bibnamefont{Tlusty}}, \bibnamefont{and}
  \bibinfo{author}{\bibfnamefont{R.}~\bibnamefont{Bar-Ziv}},
  \bibinfo{journal}{Nature Physics} \textbf{\bibinfo{volume}{2}},
  \bibinfo{pages}{743} (\bibinfo{year}{2006}).

\bibitem[{\citenamefont{Beatus et~al.}(2007)\citenamefont{Beatus, Bar-Ziv, and
  Tlusty}}]{beatus2007}
\bibinfo{author}{\bibfnamefont{T.}~\bibnamefont{Beatus}},
  \bibinfo{author}{\bibfnamefont{R.}~\bibnamefont{Bar-Ziv}}, \bibnamefont{and}
  \bibinfo{author}{\bibfnamefont{T.}~\bibnamefont{Tlusty}},
  \bibinfo{journal}{Phys. Rev. Lett.} \textbf{\bibinfo{volume}{99}},
  \bibinfo{pages}{124502} (\bibinfo{year}{2007}).

\bibitem[{\citenamefont{Derrida and Evan}(1997)}]{derridabook}
\bibinfo{author}{\bibfnamefont{B.}~\bibnamefont{Derrida}} \bibnamefont{and}
  \bibinfo{author}{\bibfnamefont{M.~R.} \bibnamefont{Evan}},
  \emph{\bibinfo{title}{Non-Equilibrium Statistical Mechanics in One
  Dimension}} (\bibinfo{publisher}{Cambridge university press},
  \bibinfo{year}{1997}).

\bibitem[{\citenamefont{Helbing}(2001)}]{helbing2001}
\bibinfo{author}{\bibfnamefont{D.}~\bibnamefont{Helbing}},
  \bibinfo{journal}{Rev. Mod. Phys.} \textbf{\bibinfo{volume}{73}},
  \bibinfo{pages}{1067} (\bibinfo{year}{2001}).

\end{thebibliography}

\end{document}